# Voltage Support Capability Analysis of Grid-Forming Inverters with Current-Limiting Control Under Asymmetrical Grid Faults

Han Zhang, *Graduate Student Member,* Rui Liu, *Graduate Student Member, IEEE*, and Yunwei (Ryan) Li, *Fellow, IEEE*

*Abstract*—**Voltage support capability is critical for grid-forming (GFM) inverters with current-limiting control (CLC) during grid faults. Despite the findings on the voltage support for symmetrical grid faults, its applicability to more common but complex asymmetrical grid faults has yet to be verified rigorously. This letter fills the gap in the voltage support capability analysis for asymmetrical grid faults by establishing and analyzing positive- and negative-sequence equivalent circuit models, where the virtual impedance is adopted to emulate various CLCs. It is discovered that matching the phase angle of the virtual impedance, emulated by the CLC, with that of the composed impedance from the capacitor to the fault location can maximize the voltage support capability of GFM inverters under asymmetrical grid faults. Rigorous theoretical analysis and experimental results verify this conclusion.**

## I. INTRODUCTION

GRID-FORMING (GFM) inverters are crucial components of the modern and futural power system. This is particularly true for a power system with a high penetration of power electronics-based devices, as the inverters possess superior voltage and frequency support capabilities as well as high robustness against weak grids [1]. Nevertheless, GFM inverters easily suffer from severe overcurrent brought by large grid disturbances, especially from grid faults, due to the limited thermal inertia of switches. Besides, grid codes require inverters to maintain the connection with the power system in case of cascaded disturbances and even blackouts [2]. Therefore, in the past decade, numerous current-limiting controls (CLCs) for GFM inverters have been developed to ride through grid faults [3].

Presently, the majority of existing current-limiting controls and research papers on GFM inverters under grid faults focus on fast current limiting [3] and transient stability enhancement [4], whereas the voltage support capability has rarely been investigated. However, despite such negligence, voltage support is essentially a crucial requirement, as it is beneficial in maintaining the system voltage stability during faults and enhancing the grid voltage recovery after fault clearance, thereby improving the system's resilience [5, 6]. Therefore, the voltage support capability should be sufficiently analyzed and enhanced for GFM inverters under grid faults.

Unfortunately, advanced voltage support enhancement control strategies and theoretical analysis of grid-following (GFL) inverters are inapplicable to GFM inverters since the inverters behave as a voltage source behind an impedance [7,

8]. Although the voltage support capability of GFM inverters with CLC is theoretically elaborated and experimentally verified for symmetrical grid faults [9], for more complex and common asymmetrical grid faults, only simulation verification of GFM inverters with a virtual impedance (VI) is developed [10]. Therefore, it is necessary to theoretically and experimentally investigate how to maximize the voltage support capability of GFM inverters with CLC under asymmetrical grid faults.

To fill the gap in the voltage support capability analysis of GFM inverters with current-limiting control under asymmetrical grid faults, unified positive- and negative-sequence equivalent circuit models are first established, where the VI is adopted to emulate various CLCs [9]. The relationships between positive- and negative-sequence currents and capacitors are derived from the equivalent circuit models, and the optimal approach to maximize the voltage support capability is obtained and verified by rigorous theoretical analysis and experimental results. It is observed that matching the phase angle of the VI with that of the composed impedance from the capacitor to the fault location can maximize the voltage support capability of GFM inverters under asymmetrical grid faults. Despite a similar conclusion to [9, 10], the significant contributions of this letter are summarized:

1) This letter first provides a comprehensive and rigorous theoretical analysis of the voltage support capability of GFM inverters under asymmetrical grid faults. The proposed analysis unifies and completes the voltage support capability analysis theory under different types of grid faults since the symmetrical grid fault can be regarded as a special case in this analysis.

2) The finding of the voltage support capability can be adopted to guide the design of futural current-limiting controls for GFM inverters under asymmetrical grid faults, providing a practical and simple approach to accomplish the maximum voltage support capability.

## II. SYSTEM DESCRIPTION AND CONTROL LOOPS

### A. System Description

The single-line diagram of a GFM inverter is depicted in Fig. 1, where $v_{dc}$, $u$, $v_C$, and $v_g$ respectively denote the DC-link voltage, the inverter terminal voltage, the capacitor voltage, and the grid voltage. $i$ and $i_o$ denote the filter current and output current. $L_f$ denotes the inverter-side filter inductance, $C_f$ denotes the filter capacitance, and $L_T$ denotes the leakage inductance of the transformer $T_0$; thus, they constitute an *LCL* filter to attenuate high-frequency ripples. $Z_{g1}$ and $Z_{g2}$ denote the equivalent grid impedance. $Z_f$ denotes the fault impedance. $f$

denotes the fault location. Lastly, $P$ and $Q$ denote the output active power and reactive power, respectively.

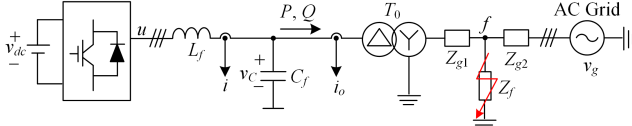

Fig. 1. Single-line diagram of a grid-connected GFM inverter.

### B. Control Loops

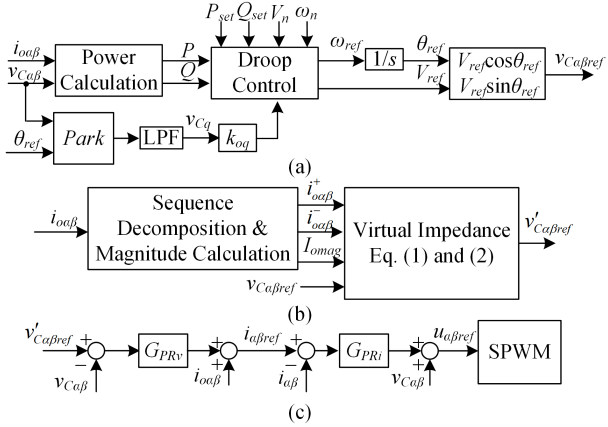

Fig. 2. Block diagrams of control loops. (a) Outer power control loop. (b) Adaptive VI. (c) Inner voltage and current control loops.

Droop control in Fig. 2(a) and the cascaded vector voltage and current control in Fig. 2(c) are utilized as the GFM scheme in this paper. Since they are classic control schemes, no further elaborations are provided in this letter. As shown in Fig. 2(a), $k_{oq}V_{Cq}$ ensures the transient stability during faults [11]. Fast current limiting is not considered in this letter since it will not affect the voltage support capability during faults, and it can be simply implemented by supplementing a circular current limiter [12] for transient overcurrent limiting.

VI can emulate different CLCs in fault analysis, simplifying and unifying the circuit analysis [9, 13]; hence, this letter adopts an adaptive VI to represent different CLCs [14], as shown in Fig. 2(b). The expression of the adaptive VI is shown as

$$R_v = X_v / n_{XR}, X_v = \begin{cases} K_X \left( I_{omag} - I_{th} \right), \text{if } I_{omag} \geq I_{th} \\ 0, & \text{if } I_{omag} < I_{th} \end{cases} \quad (1)$$

where $I_{omag}$ denotes the maximum magnitude of $i_o$ in the natural reference frame. $R_v$ and $X_v$ respectively denote the adaptive virtual resistance and reactance for current limiting during faults. $n_{XR}$ denotes the X/R ratio of the adaptive VI. $I_{th}$ is the current threshold beyond which the adaptive VI is activated, and $K_X$ is the proportion gain of the adaptive VI. With the adoption of the adaptive VI, the final capacitor voltage reference can be deduced as [15]

$$\begin{cases} v'_{Caref} = v_{Caref} - \left( R_v i_{o\alpha}^+ - X_v i_{o\beta}^+ + R_v i_{o\alpha}^- + X_v i_{o\beta}^- \right) \\ v'_{C\beta ref} = v_{C\beta ref} - \left( R_v i_{o\beta}^+ + X_v i_{o\alpha}^+ + R_v i_{o\beta}^- - X_v i_{o\alpha}^- \right) \end{cases} \quad (2)$$

where $v_{Caref}$ and $v_{C\beta ref}$ are voltage references from the droop control. $i_{o\alpha\beta}^+$ and $i_{o\alpha\beta}^-$ respectively denote the positive-sequence and negative-sequence inverter output currents, decomposed from inverter output currents $i_{o\alpha\beta}$ [16]. The sequence decomposition method is omitted here since it is fundamental and well-known.

### III. Voltage Support Capability Analysis Under Asymmetrical Grid Faults

To calculate positive-sequence and negative-sequence capacitor voltages under asymmetrical grid faults, positive- and negative-sequence equivalent circuit models of the GFM inverter are established, as shown in Fig. 3. There is no zero-sequence circuit model since there is no zero-sequence path for three-wire inverters. $\dot{v}_{Cref}^+$ denotes the positive-sequence reference voltage vector. $\dot{v}_f^+$ and $\dot{v}_f^-$ denote the positive-sequence and negative-sequence faulty voltage vectors, respectively. $\dot{v}_C^+$ and $\dot{v}_C^-$ denote the positive-sequence and negative-sequence capacitor voltage vectors, respectively. $\dot{Z}_v$ and $\dot{Z}_L$ denote the VI vector and composed impedance from the fault location to the capacitor, respectively. In this paper, $\dot{Z}_L = \dot{Z}_T + \dot{Z}_{g1}$, where $\dot{Z}_T$ denotes the transformer impedance vector.

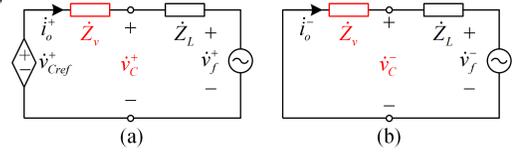

Fig. 3. Equivalent circuit models of the GFM inverter under asymmetrical grid faults. (a) Positive-sequence. (b) Negative-sequence.

As shown in Fig. 3, the positive-sequence and negative-sequence inverter output current vectors can be calculated as

$$\dot{i}_o^+ = \frac{\dot{v}_{Cref}^+ - \dot{v}_f^+}{\dot{Z}_v + \dot{Z}_L}, \dot{i}_o^- = -\frac{\dot{v}_f^-}{\dot{Z}_v + \dot{Z}_L} \quad (3)$$

The inverter output currents in the natural reference frame can be calculated as [17]

$$\dot{i}_{oa} = \dot{i}_o^+ + \dot{i}_o^-, \dot{i}_{ob} = a^2 \dot{i}_o^+ + a \dot{i}_o^-, \dot{i}_{oc} = a \dot{i}_o^+ + a^2 \dot{i}_o^- \quad (4)$$

where the operator $a = j120°$. Substituting (3) into (4), the three-phase inverter output currents can be expressed as

$$\dot{i}_{oa} = \frac{\dot{v}_{Cref}^+ - \dot{v}_f^+ - \dot{v}_f^-}{\dot{Z}_v + \dot{Z}_L} = \frac{\dot{v}_{Ma}}{\dot{Z}_v + \dot{Z}_L}$$

$$\dot{i}_{ob} = \frac{a^2 \left( \dot{v}_{Cref}^+ - \dot{v}_f^+ \right) - a \dot{v}_f^-}{\dot{Z}_v + \dot{Z}_L} = \frac{\dot{v}_{Mb}}{\dot{Z}_v + \dot{Z}_L} \quad (5)$$

$$\dot{i}_{oc} = \frac{a \left( \dot{v}_{Cref}^+ - \dot{v}_f^+ \right) - a^2 \dot{v}_f^-}{\dot{Z}_v + \dot{Z}_L} = \frac{\dot{v}_{Mc}}{\dot{Z}_v + \dot{Z}_L}$$

As shown in (5), for certain asymmetrical grid faults, $\dot{v}_{Ma}$, $\dot{v}_{Mb}$, and $\dot{v}_{Mc}$ are determined and independent of the VI and composed impedance. Due to the same denominator of three-phase currents, their magnitude relationship is also definite, and magnitudes of three-phase inverter output currents are linear to the magnitude of the sum of $\dot{Z}_v$ and $\dot{Z}_L$. Considering the maximum voltage support capability, the maximum current magnitude is limited to be the maximum allowable current magnitude $I_M$ [9]. As a result, the magnitude of the sum of $\dot{Z}_v$ and $\dot{Z}_L$ is determined, while the phase of the sum of $\dot{Z}_v$ and $\dot{Z}_L$ is variable and dependent on the X/R ratio of the VI.

As shown in Fig. 3, the positive-sequence and negative-sequence capacitor voltages can be deduced as

$$\dot{v}_C^+ = \dot{Z}_L \dot{i}_o^+ + \dot{v}_f^+ = \dot{v}_{C\_f}^+ + \dot{v}_f^+$$

$$\dot{v}_C^- = \dot{Z}_L \dot{i}_o^- + \dot{v}_f^- = \dot{v}_{C\_f}^- + \dot{v}_f^- \quad (6)$$

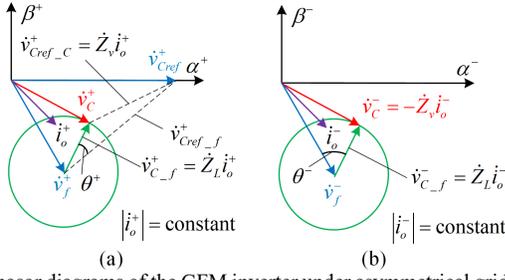

Fig. 4. Phasor diagrams of the GFM inverter under asymmetrical grid faults. (a) Positive-sequence phasor diagram. (b) Negative-sequence phasor diagram.

To analyze the relationship among the derived vectors in this paper, the positive-sequence and negative-sequence phasor diagrams of the GFM inverter under asymmetrical grid faults are respectively illustrated in Fig. 4(a) and (b). Since the magnitude of the sum of $\dot{Z}_v$ and $\dot{Z}_L$ is determined, as shown in (3), magnitudes of the positive-sequence and negative-sequence currents are determined. Therefore, as shown in Fig. 4, $|\dot{i}_o^+|$ and $|\dot{i}_o^-|$ are constants. Besides, $\dot{Z}_L$ is also determined; hence, $\dot{v}_{C\_f}^+$ and $\dot{v}_{C\_f}^-$ are circles with constant radii of $|\dot{Z}_L \dot{i}_o^+|$ and $|\dot{Z}_L \dot{i}_o^-|$, respectively.

Maintaining the pre-fault voltages to the maximum extent under faults without causing overcurrent presents the maximum voltage support capability for GFM inverters [18]. In other words, the variation of capacitor voltages under grid faults should be minimized to the largest extent. As shown in Fig. 4(a) and (b), the positive-sequence capacitor voltage variation is $\dot{v}_{Cref\_C}^+$, and the negative-sequence capacitor voltage variation is $\dot{v}_C^-$. The appropriate X/R ratio of the VI is investigated in this paper to simultaneously minimize the two capacitor voltage variations.

According to Fig. 4(a) and (b), magnitudes of $\dot{v}_{Cref\_C}^+$ and $\dot{v}_C^-$ can be calculated as

$$\left| \dot{v}_{Cref\_C}^+ \right| = \sqrt{\left| \dot{v}_{C\_f}^+ \right|^2 + \left| \dot{v}_{Cref\_f}^+ \right|^2 - 2 \left| \dot{v}_{C\_f}^+ \right| \left| \dot{v}_{Cref\_f}^+ \right| \cos \theta^+}$$
$$\left| \dot{v}_C^- \right| = \sqrt{\left| \dot{v}_f^- \right|^2 + \left| \dot{v}_{C\_f}^- \right|^2 - 2 \left| \dot{v}_f^- \right| \left| \dot{v}_{C\_f}^- \right| \cos \theta^-}$$

(7)

where $\theta^+$ and $\theta^-$ denote the phase angle differences between $\dot{v}_{C\_f}^+$ and $\dot{v}_{Cref\_C}^+$ and $\dot{v}_{C\_f}^-$ and $\dot{v}_f^-$, respectively. When the grid fault is determined, $|\dot{v}_{C\_f}^+|$, $|\dot{v}_{Cref\_f}^+|$, $|\dot{v}_{C\_f}^-|$, and $|\dot{v}_f^-|$ are constant values as previously analyzed. Therefore, when $\theta^+ = 0$ and $\theta^- = 0$, $|\dot{v}_{Cref\_C}^+|$ and $|\dot{v}_C^-|$ are respectively minimized to the lowest values, i.e., the positive-sequence and negative-sequence voltage support capabilities are maximized.

As shown in Fig. 4, $\theta^+ = 0$ and $\theta^- = 0$ reveal that the phase angle of $\dot{Z}_v$ is equal to the phase angle of $\dot{Z}_L$. Therefore, when the phase angle of the VI is equal to that of the composed impedance from the inverter to the fault location, the voltage support capability of GFM inverters under asymmetrical grid faults can be maximized.

The symmetrical grid faults can be regarded as a special case in the above analysis by neglecting the negative-sequence equivalent circuit model; thus, the analysis and conclusion for asymmetrical grid faults can also be applied to symmetrical grid faults, which is verified in [9].

## IV. Experimental Verification

The experiment is carried out on a platform with the same topology as in Fig. 1. The grid is simulated by a MX30 power supply, the DC voltage is provided by a Keysight power supply, and the dSPACE MicroLabBox DS1202 is adopted as the digital controller. The fault resistance is cut in and off the circuit via the solid-state relay SSR3-D48100ZK, and a line-to-line (LL) fault is developed to test the theoretical analysis. All critical experimental parameters are listed in Table I.

TABLE I
EXPERIMENTAL SYSTEM PARAMETERS

| Symbol | Description | Value |
|---|---|---|
| $S_n$ | Rated power | 500VA |
| $V_{dc}/V_g$ | DC Voltage/ Grid voltage (LL, RMS) | 200V/104V |
| $L_f/C_f$ | Filter inductance/capacitance | 3mH/30μF |
| $X_T$ | Inductance of transformer $T_0$ | 0.028pu |
| $Z_{g1}/Z_{g2}$ | Equivalent grid impedance 1 and 2 | 3mH; 5mH/1Ω |
| $Z_f$ | Fault impedance | 6.8Ω |
| $P_{set}/Q_{set}$ | Setting active/reactive power value | 1p.u./0p.u. |
| $\omega_n$ | Nominal angular frequency | 314rad/s |
| $V_n$ | Nominal voltage magnitude | 84.85V |
| $I_M$ | Maximum allowable current magnitude | 1.5p.u. |
| $T_s$ | Sampling period | 100μs |

The maximum allowable current magnitude $I_M$ is set as 1.5p.u. [9, 12]. Similar to [9], for a fair comparison of the voltage support capability, the maximum current magnitude is limited at $I_M$ for the VI. Fig. 5 illustrates the experimental results of the capacitor voltages, inverter output currents, positive-sequence capacitor voltage magnitude, and negative-sequence capacitor voltage magnitude under an LL fault with two types of VI. Both VIs realize the current limiting, and the maximum current magnitudes during faults are both limited at $I_M$. Nevertheless, comparing Fig. 5(b) and (a), the inductive VI can accomplish a higher voltage support capability during a fault than the resistive VI. Particularly, the positive-sequence capacitor voltage magnitude with the inductive VI is 69.4V, which is higher than the 62.7V with the resistive VI, and the negative-sequence capacitor voltage magnitude is 10.6V with the inductive VI, which is lower than the 13.6V with the resistance VI. The experimental results comply with the theoretical analysis since the composed impedance from the capacitor to the fault location is inductive.

To further verify that matching the VI with the composed impedance can maximize the voltage support capability under asymmetrical grid faults, positive-sequence and negative-sequence capacitor voltage magnitudes during an LL fault are compared in Fig. 6 for the phase angles from 0° to 90° of the VI. As shown in Fig. 6, as the phase angle of the VI increases from 0° to 75°, the positive-sequence capacitor voltage magnitude increases and the negative-sequence capacitor voltage decreases as the VI becomes more inductive, gradually matching with the composed impedance from the capacitor to the fault location. When the phase angle of the VI is 90°(purely inductive), the voltage support capability slightly reduces, which is reasonable since the composed impedance is not purely inductive due to parasitic resistance.

Therefore, the above experimental results can sufficiently verify that matching the phase angle of the VI with that of the composed impedance can significantly enhance the voltage support capability of GFM inverters under asymmetrical grid faults.

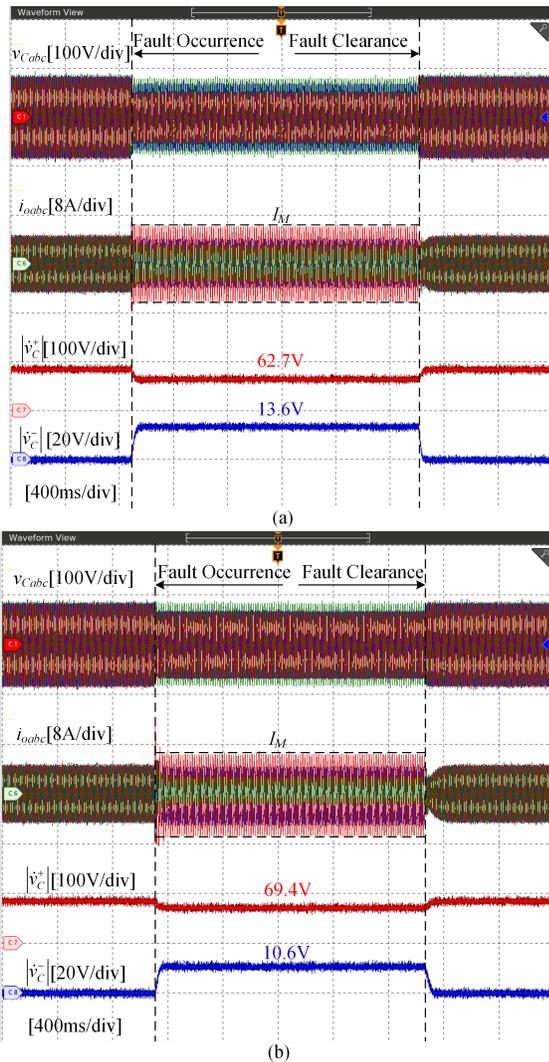

Fig. 5. Experimental results of capacitor voltages, inverter output currents, positive-sequence capacitor voltage magnitude, and negative-sequence capacitor voltage magnitude under an LL fault with different VIs. (a) Resistive VI (phase angle is 0°). (b) Inductive VI (phase angle is 75°).

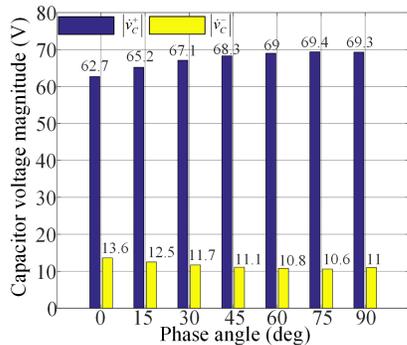

Fig. 6. Voltage support capability comparsion under an LL fualt for different phase angles of the VI.

## VI. CONCLUSION

Notwithstanding the derivation of the voltage support analysis under symmetrical grid faults, the analysis under asymmetrical grid faults is more testing yet beneficial since the faults are more sophisticated but prevalent. Rigorous analysis and experiments substantiate that matching the phase angle of the virtual impedance, emulated by the current-limiting control, with that of the composed impedance from the capacitor to the fault location can optimize the voltage support capability of GFM inverters under asymmetrical grid faults, contributing to and unifying the voltage support analysis under symmetrical and asymmetrical grid faults.